\documentclass[]{aa}

\usepackage{psfig}

\def\etal{et\,al.}
\def\msun{M$_{\odot}$}

\def\degs{\ifmmode ^{\circ}\else$^{\circ}$\fi}
\def\amin{\ifmmode ^{\prime}\else$^{\prime}$\fi}
\def\asec{\ifmmode ^{\prime\prime}\else$^{\prime\prime}$\fi}
\def\fss{\hbox{$.\!\!^{\rm s}$}}        
\def\farcs{\hbox{$.\!\!^{\prime\prime}$}}  
\def\farcm{\hbox{$.\mkern-4mu^\prime$}}    
\newbox\grsign \setbox\grsign=\hbox{$>$}
\newdimen\grdimen \grdimen=\ht\grsign
\newbox\laxbox \newbox\gaxbox
\setbox\gaxbox=\hbox{\raise.5ex\hbox{$>$}\llap
     {\lower.5ex\hbox{$\sim$}}}\ht1=\grdimen\dp1=0pt
\setbox\laxbox=\hbox{\raise.5ex\hbox{$<$}\llap
     {\lower.5ex\hbox{$\sim$}}}\ht2=\grdimen\dp2=0pt
\def\gax{\mathrel{\copy\gaxbox}}

\def\h{$^{\rm h}$}\def\m{$^{\rm m}$}

\def\v7{V751 Cyg}
\def\rxj0513{RX\,J0513.9--6951}

\begin{document}

   \title{BZ Cam during its 1999/2000 optical low state}

   \titlerunning{BZ Cam's 1999/2000 low state}

   \author{J. Greiner\inst{1}, G. Tovmassian\inst{2}, M. Orio\inst{3,4},
      H. Lehmann\inst{5},
      V. Chavushyan\inst{6}, A. Rau\inst{1}, R. Schwarz\inst{1},
      R. Casalegno\inst{3,4},
      R.-D. Scholz\inst{1}}

   \authorrunning{Greiner \etal}

   \offprints{J. Greiner, jgreiner@aip.de}

   \institute{Astrophysical Institute
             Potsdam, An der Sternwarte 16, D-14482 Potsdam, Germany
             \and
             OAN, Instituto de Astronom\'{\i}a, UNAM, AP 877, 22860 Ensenada,
                 M\'{e}xico
             \and
             Osservatorio Astronomico di Torino, Strada Osservatorio 20,
            I-10125 Pino Torinese (TO), Italy
             \and
            Dept. of Astronomy, Univ. Wisconsin, 1150 University Ave.,
             Madison, WI 53706, USA
             \and
             Th\"uringer Landessternwarte, Sternwarte 5, D-07778 Tautenburg,
             Germany
             \and
             Instituto Nacional de Astrofisica, Optica y Electronica (INAOE),
              Aptdo. Postal 51 y 216, 72000 Puebla, Pue., Mexico
             }

   \date{Received 11 April 2001; accepted 20 July 2001}

\abstract{
  We report optical observations of the VY Scl star BZ Cam during
  its previous optical low state in 1999/2000.
  We find drastic variations in the line profiles.
  Narrow-band imaging observations show that its nebula extends
  farther than previously known and seems to be composed of two components.
  We determine the \ion{[O}{III]} line intensity of BZ Cam's nebula
   to 4.8$\times$10$^{-13}$ erg/cm$^2$/s.
  We discover a proper motion  of BZ Cam of 25$\pm$2 mas/yr which
  together with the systemic radial velocity yields a space velocity
  of 125 km/s.
  We re-interpret the nebula as being photo-ionized by hypothesized
  transient, luminous, supersoft X-ray emission during optical low states,
  and shaped by the transverse motion of BZ Cam.
    }

    \maketitle

\section{Introduction}

BZ Cam is a binary system with a period of 221 min (Patterson \etal\ 1996).
It is thought to contain
an accreting white dwarf and a 0.3--0.4 \msun\ main-sequence donor
(Lu \& Hutchings 1985).
BZ Cam belongs to the group of variable stars called VY Scl stars, or
anti-dwarf novae, due to its occasional drop in brightness. Most of the time
it spends at around V = 12.0-12.5 mag ($\pm$ 0.2 mag), but during low states
it is as low as V = 14.3 mag. Only one previous optical low state is known,
which occurred in 1928 (Garnavich \& Szkody 1988).

BZ Cam is surrounded by a faint emission nebula (Ellis \etal\ 1984)
which has a bow-shock-like structure (Krautter \etal\ 1987,
Hollis \etal\ 1992). This nebula is also detected at radio frequencies
(Hollis \etal\ 1992), implying a $\sim$35 cm$^{-3}$ density in the \ion{H}{II}
recombination region (assuming an electron temperature of 10$^4$ K).
Based on the optical emission line ratios these authors argue
that photoionization can not alone account for the
excitation of the nebula, and that  shock wave heating seems to contribute.
Based on IUE data BZ Cam was found to exhibit a wind
(Hollis \etal\ 1992), which was recently also detected in the optical 
as well as to display rapid variability
(Ringwald \& Naylor 1998), rare among canonical CVs.

Supersoft X-ray binaries (SSB) were established as new class of astronomical
objects during the early 90ies  (Greiner 2000).
They contain a white dwarf (WD), accreting  mass at
rates high enough to allow quasi-steady nuclear surface burning
(van den Heuvel \etal\ 1992). Their luminosities are of order of
$L_{bol} \sim 10^{36}-10^{38}$ erg s$^{-1}$, with typical temperatures
of 30--50 eV.
Two SSBs have particular properties which are worth mentioning in this
context:
(i) CAL 83: It is the only SSB (among a dozen SSBs searched)
  which is surrounded by a distinct nebula (Pakull \& Motch 1989, Remillard 
  \etal\ 1995) caused by ionization of the surrounding interstellar medium
  by the luminous X-ray radiation (Rappaport 1994).
(ii) \rxj0513: It shows quasi-periodic optical low states
  (Southwell \etal\ 1996) during which supersoft X-ray emission is ``on''
  (Schaeidt \etal\ 1993).
Among the ``classical'' CVs two systems have recently been shown to
exhibit transient, supersoft X-ray emission, both during periods of
optical low states: the VY Scl star V751 Cyg (Greiner \etal\ 1999)
and  V Sge (Greiner \& Teeseling 1998).

\begin{figure*}
 \vbox{\psfig{figure=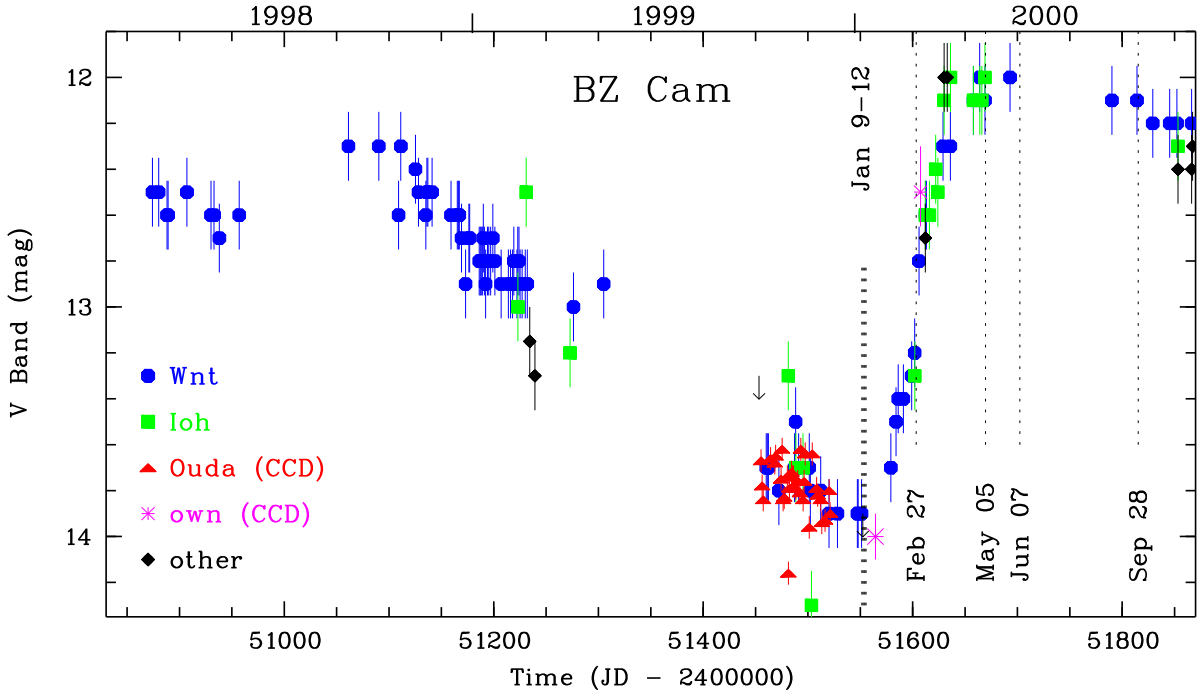,width=16.cm,angle=270,%
   bbllx=4.3cm,bblly=2.9cm,bburx=18.5cm,bbury=27.2cm,clip=}}\par
 \caption[fc]{Optical light curve of BZ Cam with most of the data taken
           from VSNET. The times of optical spectroscopy and narrow-filter
           imaging are marked by vertical dotted lines.
 \label{lc}}
\end{figure*}

BZ Cam recently entered a very rare optical low state.
Fig. \ref{lc} shows the optical light curve of BZ Cam over the last
2 years covering the recent optical low state. Due to the prolonged
intermediate brightness state at around V $\sim$12.5--13.2 mag since
at least 1997 it is difficult to
accurately establish the beginning of this recent low state, but a
reasonable estimate is after JD = 2451300 (see Fig. \ref{lc}).
BZ Cam moved out of its low state around Feb./Mar. 2000 (starting
around JD = 2451580; note the seemingly slow fading, but rapid rise).

We report here on our observations triggered by this rare optical
low state.

\section{Observations and Interpretation}

\begin{table}
\caption{Log of Observations}
\vspace{-0.2cm}
\label{log}
\begin{tabular}{llllr}
      \hline
      \noalign{\smallskip}
      ~~Teles-     & ~Date  &  Filter/ & D$^{(2)}$ & T$_{\rm Int}$ \\
   ~~cope$^{(1)}$  & ~(2000) &  Wavelength  & (hrs) & (sec) \\
      \noalign{\smallskip}
      \hline
      \noalign{\smallskip}
 INAOE 2.1m  &  Jan 9--12 & 4000--7500 & 1.0$^{(3)}$ & 1500 \\
 AIP 0.7m    &  Jan 21 & white      & 1.7 & 10 \\
 AIP 0.7m    &  Feb 02 & V      & 0.5 & 30 \\
 Tbg 2.0m    &  Feb 27 & 3500--9500 & 1.9 & 1800 \\
 INAOE 2.1m  &  Feb 26--28 & 4000--7500 & 1.0$^{(3)}$ & 1800 \\
 OAN 1.0m  &  Mar 04 & V/R & 0.5 & 60 \\
 INAOE 2.1m  &  May 05 & 4000--7500 & 4.0  & 1500 \\
 Tbg 2.0m    &  Jun 7/8 & 4000--8500 & 1.9 & 1800 \\
 WIYN 3.5m   &  Sep 28 & \,[OIII], H$\alpha$ & $-\!\!-$ & 600 \\
 \noalign{\smallskip}
 \hline
 \noalign{\smallskip}
 \end{tabular}

\noindent{
  $^{(1)}$ The abbreviations mean:
  AIP = Astrophysical Institute Potsdam, Germany;
  OAN = Tonantzintla, Mexico \\
  INAOE = 2.1m at Cananea Observatory, 1.0m at Tonantzintla, Mexico;
  Tbg = Tautenburg Observatory, Germany;
  WIYN = WIYN Telescope, Kitt Peak, USA \\
 $^{(2)}$ Duration of the temporal photometric coverage. \\
 $^{(3)}$ 1 hr in each night.
  }
 \end{table}

We acquired  photometric and spectroscopic observations (see Tab. \ref{log}
for a log) in the optical low state,  during the rise out of the optical
low-state and in the subsequent optical high-state.
The optical light curve (including data compiled from VSNET) with the times
of our observations marked is shown in Fig. \ref{lc}.

Low resolution spectra  of BZ\,Cam were obtained  at the 2.1\,m telescope
at Cananea, M\'exico in winter  of 2000. The LFOSC spectrograph
(Zickgraf \etal\ 1997) was deployed in mid-January (8--11)
observations to cover the $\lambda$\,4000--7500\,\AA\, wavelength range with
$\approx$\,13\,\AA\, FWHM resolution. A 3 arcsec wide long slit, as
projected on the sky, was used at the entrance of this multi-object 
spectrograph. Later in the month (27-29 January) the object
was again observed using the same telescope, but with the B\&Ch spectrograph
instead of the LFOSC. The resolution  was slightly better than $\approx$10\AA,
and the coverage of wavelengths was about the same.
Again, the B\&Ch spectrograph with the same telescope and same settings was
used in May, when BZ Cam was back to its high state, to acquire a few
spectra. Spectrophotometric standard stars were observed at each
night in order to provide flux calibration. A He-Ar  arc lamp was utilized in
order to calibrate the spectra for wavelengths. IRAF standard procedures
of long slit spectroscopy were used for data processing.

The photometric measurements of the object  in the $V$ and $R$ bands were done
at the 1.0\,m telescope of OAN in Tonantzintla, M\'exico on March 4.

Observations with the 3.5\,m WIYN telescope at Kitt Peak (USA), 
operated by the University of Wisconsin, Indiana University, 
Yale University, and the National Optical Astronomy Observatories,
were done using a mosaic of four 2K$\times$2K CCDs.

Spectroscopic observations at Tautenburg were done using the newly
developed spectrograph for the Nasmyth-focus of the 2.0\,m Schmidt telescope.
Grisms with 200 \AA/mm and 100 \AA/mm were used, respectively.
Reduction of the spectra was performed using standard MIDAS routines from the
long-slit package. Wavelength calibration was done using night sky lines.

Photometric observations at the 0.7m telescope at AIP employed a 1K$\times$1K
TEK CCD camera. In the Cassegrain focus the 24 $\mu$m pixel size corresponds
to a plate scale of 0\farcs5/pixel, completely sufficient for the
typically bad seeing (2--4\asec). Initial basic reduction was done using
standard MIDAS programmes, while the photometry was done with {\sc
DoPhot} (Mateo \& Schechter 1989).

\begin{figure}[ht]
  \psfig{figure=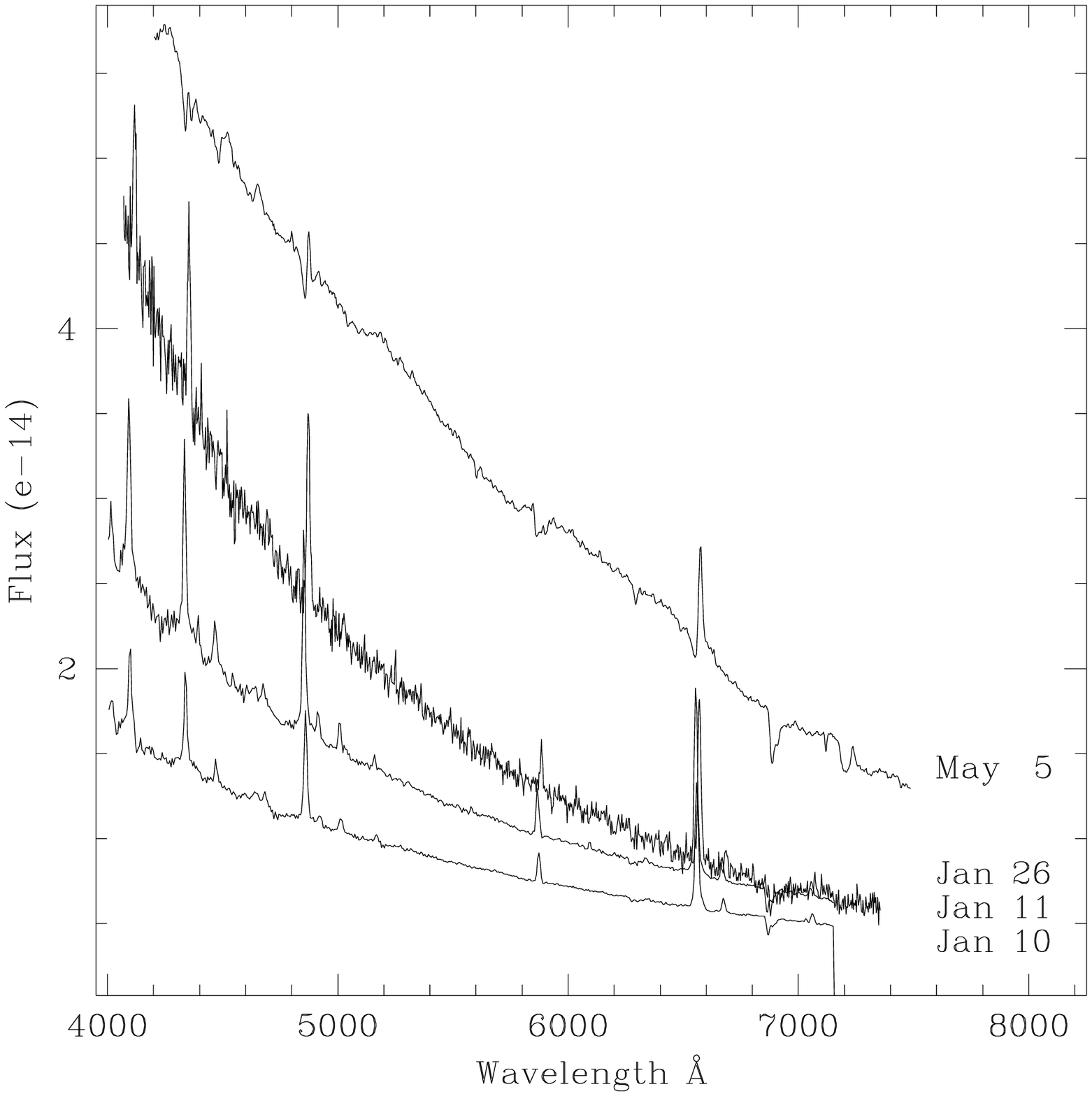,width=9.2cm}\par
  \psfig{figure=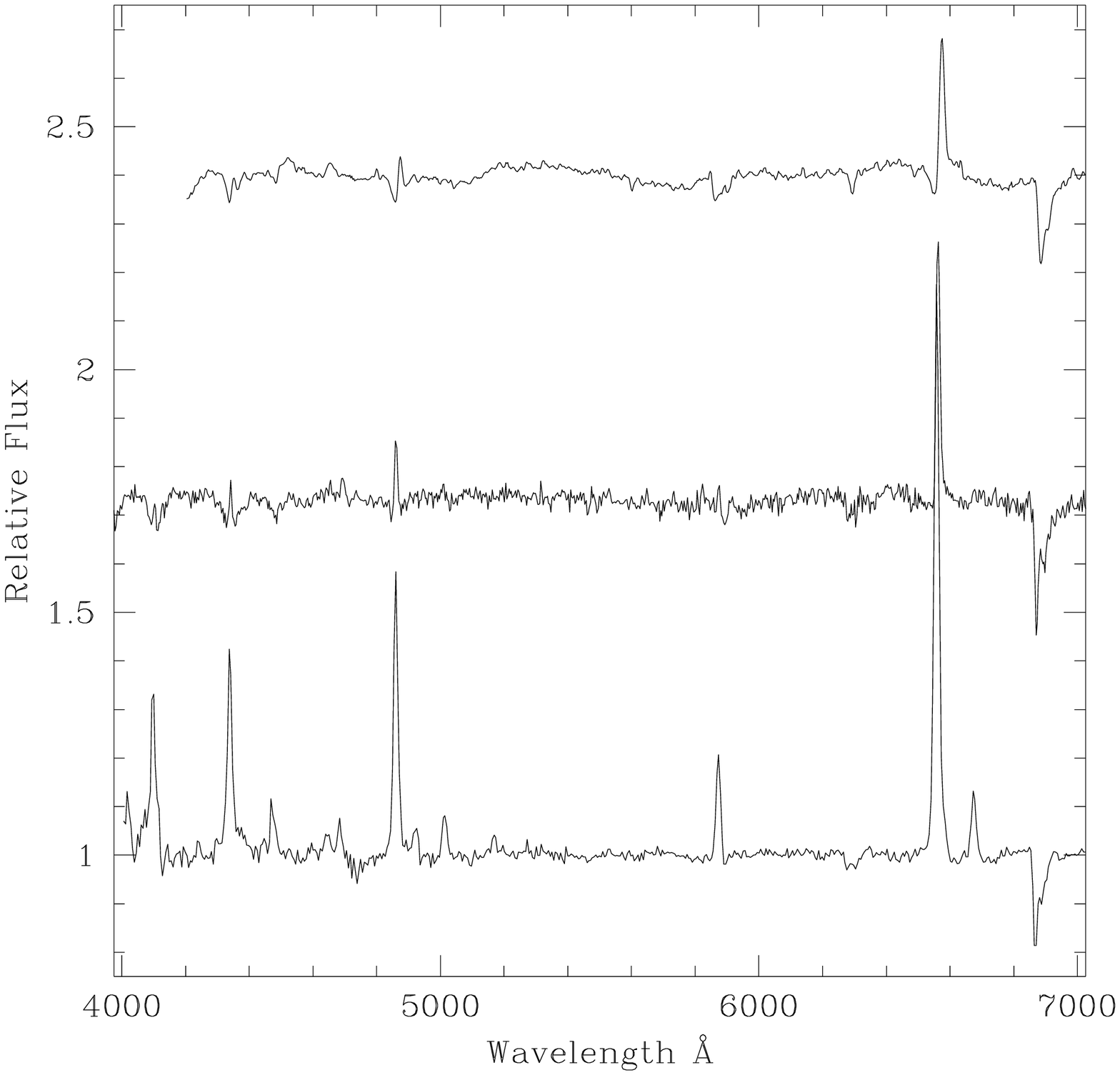,width=9.2cm}\par
 \vspace{-0.3cm}
 \caption[lst]{{\bf Top:} BZ Cam spectra in 2000: 
   low state (Jan. 10+11), beginning of the rise out of the low state (Jan. 26)
    and high-state (May 5).
    The flux is given in units of 10$^{-14}$ erg/cm$^2$/s/\AA.
    Note the different slopes and emission line strengths.
    {\bf Bottom:} Blow-up of rectified spectra of the high, intermediate
    and low state (from top to bottom). P Cygni profiles are present only 
    during the high-state, while in the intermediate state the emission
    lines appear on top of broad absorption lines.
 \label{spec}}
\end{figure}

\subsection{Low-/high-state spectra}

Sample spectra are shown in Fig. \ref{spec}, and emission line measurements
during the optical low-state are presented in Tab. \ref{linelow}
(note that during the high state the lines vary very rapidly; see e.g.
Ringwald \& Naylor 1998, so we do not present emission line ratios).
Several spectral differences are immediately recognized from the
examination of the spectra in different brightness states:
\begin{itemize}
\item\vspace{-0.2cm}
 The spectral slope is much flatter during the optical low state.
 Assuming that the spectrum is a composite of the contributions from the
 accretion disk and a hotter white dwarf, this changing slope
 suggests that the hot component gets even hotter (e.g. by reduction in size
 at constant bolometric luminosity), thus shifting
 the Wien tail out of the optical bandpass.
\item All emission lines strengthen with decreasing brightness.
\item The P Cyg profiles of (Balmer and \ion{He}{I}) emission lines
(indicative of a wind) show up only in the high state, but
are absent during the optical low  and intermediate state;
\item The \ion{He}{II} emission is absent in the high state, is visible in the
 intermediate state, and strengthens further in the low state.
\item
 With H$\gamma$ being stronger than H$\beta$ in the low state, or more 
 generally, with the Balmer line ratios not following the canonical 
 intensity ratios
 in close binaries, it seems plausible that these lines are emitted on top of
 a (sometimes broader) Balmer absorption system. This is particularly obvious
 in the intermediate state where the Balmer absorption is much broader
 than the emission system.
\end{itemize}

As was shown by Ringwald \& Naylor (1998) the emission line profiles and
their variability during the optical high state confirm the close binary
nature of BZ Cam,  but yet the profiles as well as the variability
differ significantly from ordinary CVs.
Also, the rapidly variable wind is exceptional among CVs.
This could be due to the accretion rate being very high
during the optical high state, leading possibly even to a more
spherical-like accretion geometry. One could speculate on whether or
not this may lead to the ignition of H burning. Since then the white dwarf
would burn more matter than the companion transfers into the disk, 
it could possibly empty the disk, and thus lead to the consequent onset
of an optical low state.

The non-canonical intensity ratios of the Balmer emission lines
during the optical low state (Tab. \ref{linelow})
is most probably due to a Balmer absorption system which is obvious during
the intermediate state, but possibly has not completely vanished during
the other states. We mention here that
 it is a general feature in short-period SSBs
 that e.g H$\gamma$ and H$\delta$ are clearly seen in absorption while
 H$\beta$ and H$\alpha$ are progressively filled.
Certainly, this effect is much more drastic in SSBs, but it could potentially
serve as an explanation of the emission line ratios in BZ Cam as well.

Another noteworthy property is the rather small intensity of the \ion{He}{II}
emission. This is in contrast to the fact that in SSB the \ion{He}{II}
emission line is usually the strongest line, or at least stronger than 
H$\beta$. However, this can be understood in terms of different temperatures of
the white dwarf and correspondingly different ionizing flux for \ion{He}{II}.
The mass of the white dwarf in BZ Cam is certainly smaller than 1 \msun,
and more probably in the range of 0.4--0.7 \msun\ (Lu \& Hutchings 1985).
In contrast, white dwarf masses in SSB are thought to be $\gax$1 \msun,
and consequently the effective temperatures are higher for SSB white dwarfs
as compared to BZ Cam. While this difference has little effect on the number
of ionizing photons for hydrogen ($<$912 \AA), it has a drastic effect
on the number of ionizing photons for \ion{He}{II} ($<$228 \AA).

\begin{table}
\caption{Relative emission-line fluxes of BZ Cam during the optical
low-state}
 \label{linelow}
 \vspace{-0.2cm}
 \begin{tabular}{ll}
      \hline
      \noalign{\smallskip}
      Spectral Line   & Relative Flux   \\
      \noalign{\smallskip}
      \hline
      \noalign{\smallskip}
  H$\gamma$ & 0.81-0.91   \\
  He{\small{I}}$\lambda4471$  & 0.21-0.34  \\
  H$\beta$  & 0.78-0.85   \\
  H$\alpha$  & 1.0   \\
 \noalign{\smallskip}
 \hline
 \end{tabular}
 \end{table}

\begin{table}[thb]
\caption{Positions of BZ Cam measured at different epochs}
\vspace{-0.2cm}
\begin{tabular}{rrcl}
\hline
\noalign{\smallskip}
\multispan{2}{\hfil $\alpha,\delta $(2000) \hfil}&epoch&source\\
\noalign{\smallskip}
\hline
\noalign{\smallskip}
06\h\ 29\m\ 33\fss940 & $+$71\degr\ 04\amin\ 38\farcs79 & 1897.994 & AC2000 \\
06\h\ 29\m\ 34\fss100 & $+$71\degr\ 04\amin\ 37\farcs67 & 1953.113 & POSS1-E \\
06\h\ 29\m\ 33\fss844 & $+$71\degr\ 04\amin\ 37\farcs65 & 1954.099 & POSS1-E \\
06\h\ 29\m\ 34\fss004 & $+$71\degr\ 04\amin\ 36\farcs69 & 1983.847 & POSS-V \\
06\h\ 29\m\ 33\fss992 & $+$71\degr\ 04\amin\ 37\farcs00 & 1983.847 & POSS-V \\
06\h\ 29\m\ 34\fss001 & $+$71\degr\ 04\amin\ 34\farcs28 & 1986.160 &FONAC$^*$\\
06\h\ 29\m\ 34\fss006 & $+$71\degr\ 04\amin\ 36\farcs68 & 1989.971 & POSS2-R \\
06\h\ 29\m\ 33\fss937 & $+$71\degr\ 04\amin\ 36\farcs20 & 1996.050 & POSS2-B \\
06\h\ 29\m\ 33\fss950 & $+$71\degr\ 04\amin\ 37\farcs10 & 1996.670 & HST$^*$ \\
06\h\ 29\m\ 33\fss997 & $+$71\degr\ 04\amin\ 36\farcs45 & 1997.184 & POSS2-R \\
06\h\ 29\m\ 34\fss122 & $+$71\degr\ 04\amin\ 36\farcs34 & 1997.850 & POSS2-B \\
\noalign{\smallskip}
\hline
\end{tabular}
\label{adbzcam}
\smallskip

$^*$ -- not used in final proper motion solution.
\end{table}

\subsection{Proper motion of BZ Cam}

In order to estimate the proper motion of BZ Cam, we looked for
Digitized Sky Survey (DSS) data using the plate finder service at the
Space Telescope Science Institute. On the 8 Palomar Schmidt plates
with a time baseline of more than 40 years (see Tab. \ref{adbzcam})
found in the DSS we measured the position of BZ Cam using the
plate constants provided with the FITS images and the ESO Skycat tool.
As one can see in Tab. \ref{pmbzcam}, we obtain already a clear proper
motion in negative $\delta$-direction using only these Schmidt plates
measurements, whereas in $\alpha$-direction the
proper motion is smaller than its relatively large error.

Searching the VizieR database in Strasbourg for more independent
measurements of the position of BZ Cam, we were lucky to find an
early epoch from the Astrographic Catalogue AC2000 (Urban \etal\ 1997,
Urban \etal\ 1998). In addition, there are two other positions from
the HST archive and from the FONAC catalogue (Kislyuk \etal\ 1999), both
with similar epochs as with the POSS2 data (see Tab. \ref{adbzcam}).

Combining the POSS measurements with the AC2000 position the
resulting proper motion becomes more accurate in both directions,
with $\mu_{\delta}$ reaching Hipparcos-like accuracy. The error
in $\mu_{\alpha}\cos{\delta}$ remains twice as large, mainly due
to the very different $\alpha$ positions measured on the POSS1 plates.
When the additional positions from the HST and FONAC are included,
the errors of $\mu_{\alpha}\cos{\delta}$ remain at the same level
whereas those of $\mu_{\delta}$ increase, especially after including
the FONAC position. As can be seen from Tab. \ref{adbzcam}, the
$\delta$ value from FONAC is a clear outlayer compared to all other
values. This is also the reason, why the proper motion of BZ Cam
given in the FONAC catalogue (as determined from only two positions -
from the Astrographic Catalogue and from one second epoch
plate measurement of the Kiev wide-angle astrograph) is
distinctly different in $\mu_{\delta}$ (see also Tab. \ref{pmbzcam}).
The errors of the FONAC proper motion correspond to the catalogue
precision given in Kislyuk \etal\ (1999).

For our further analysis we adopted the proper motion solution obtained
from 8 POSS plates and the AC2000:
$\mu_{\alpha}\cos{\delta}$ = 3.9$\pm$4.1 mas/yr and
$\mu_{\delta}$ = --25.1$\pm$1.8 mas/yr.
 This solution is in good agreement
with other solutions (except the FONAC $\mu_{\delta}$) and shows the
smallest total proper motion error.

\begin{table}
\caption{Proper motion solutions}
\label{pmbzcam}
\begin{tabular}{lrr}
\hline
\noalign{\smallskip}
positions used in solution & $\mu_{\alpha}\cos{\delta}$ & $\mu_{\delta}$ \\
                           & \multispan{2}{\hfil [mas/yr] \hfil} \\
\noalign{\smallskip}
\hline
8$\times$POSS                        & $+$5.1$\pm$9.2  & $-$29.9$\pm$2.9
\\
8$\times$POSS+AC2000 $^{\#}$       & $+$3.9$\pm$4.1  & $-$25.1$\pm$1.8 \\
8$\times$POSS+AC2000+HST         & $+$3.2$\pm$4.2  & $-$23.6$\pm$2.7 \\
8$\times$POSS+AC2000+HST+FONAC$\!\!$ & $+$3.2$\pm$3.9  & $-$26.3$\pm$8.6 \\
\hline
FONAC (Kislyuk \etal\ 1999) & $+$3.7$\pm$3.0 & $-$42.6$\pm$3.0\\
\noalign{\smallskip}
\hline
\end{tabular}

\smallskip
$^{\#}$ -- finally adopted proper motion solution in this paper.
\bigskip
\end{table}

\begin{figure}
  \psfig{figure=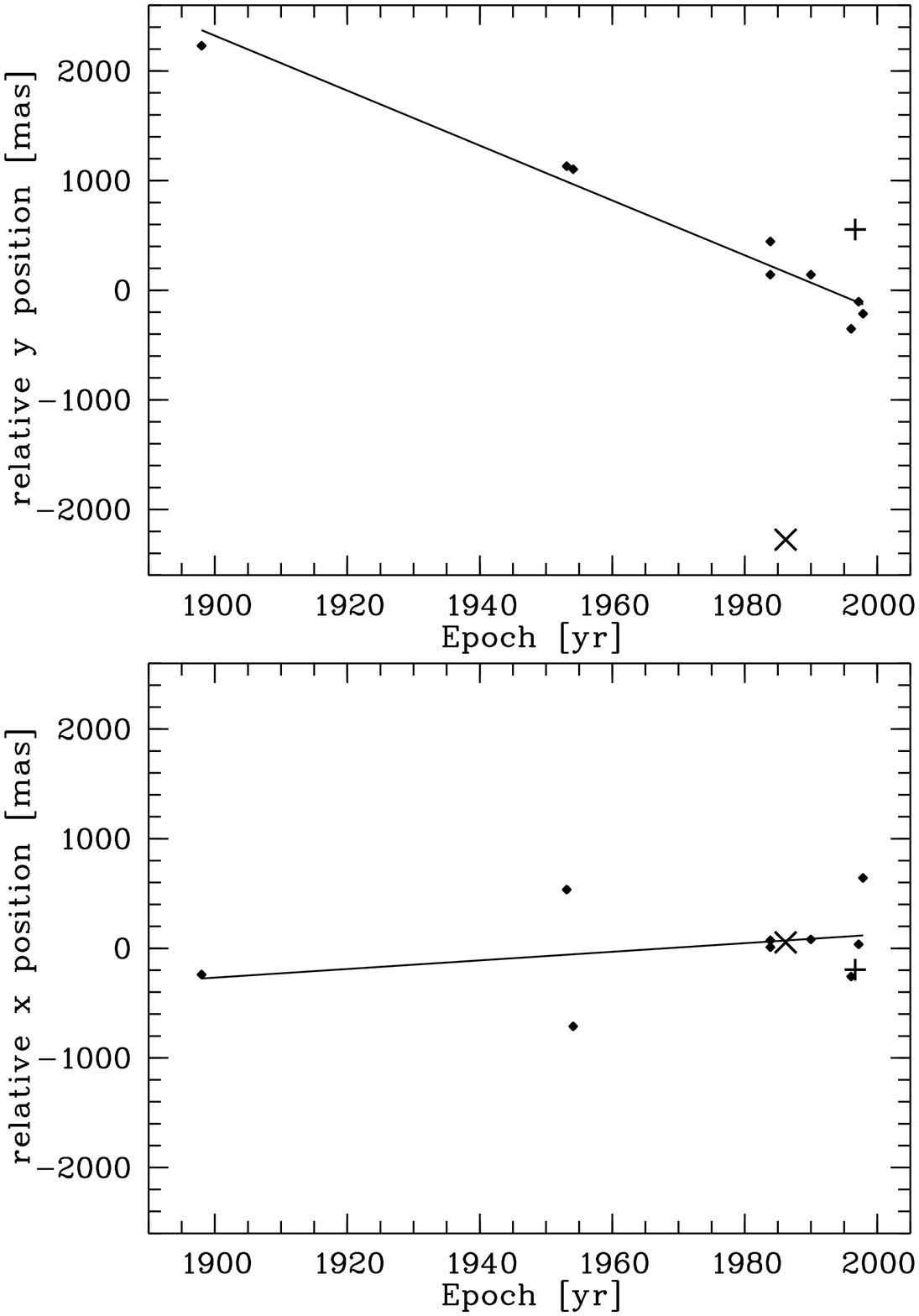,width=8.2cm,%
     bbllx=1.cm,bblly=1.7cm,bburx=19.cm,bbury=27.35cm,clip=}\par
  \caption[propmo]{The proper motion of BZ Cam obtained from measurements over
   a 100 years time baseline. The AC2000 position from 1897 and 8
   positions as measured on POSS plates between 1953 and 1997
   are shown by dots. The lines represent
   the proper motion fit using these data. The data not used in this
   finally adopted proper motion solution are shown by other symbols:
   FONAC ($\times$) and HST ($+$) (see also Tabs. \ref{adbzcam}
   and~\ref{pmbzcam}).
   \label{propmo}}
\end{figure}

At a distance of 830$\pm$160 pc (Naylor \& Ringwald 1998) this
corresponds
to a transverse velocity of 100$\pm$20 km (solely towards the South),
where most of the uncertainty stems from the distance error.
Adding a systemic radial velocity component of about --75 km/s
(Lu \& Hutchings 1985, Patterson \etal\ 1996) we derive a space velocity 
of BZ Cam of 125 km/s.

\begin{figure*}[ht]
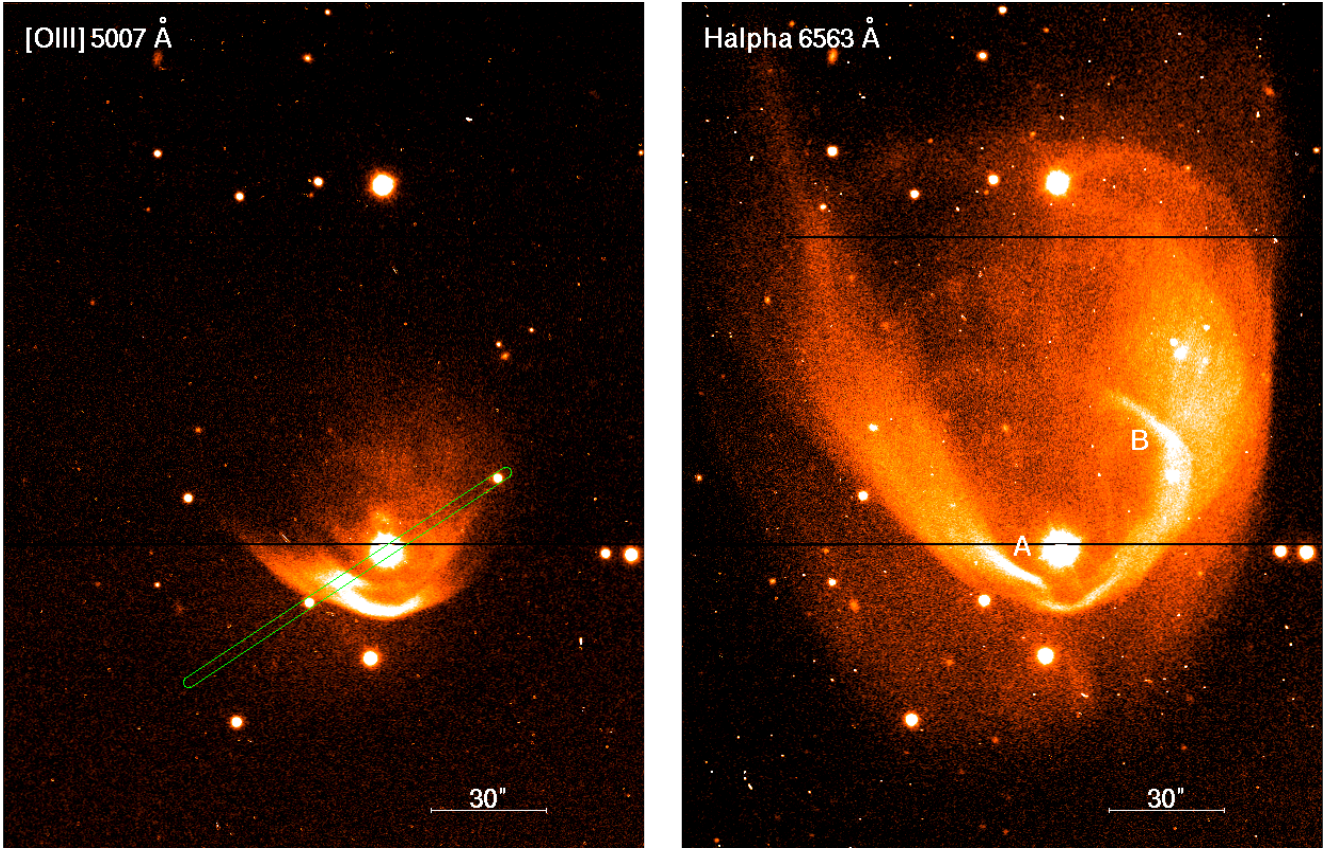

 \begin{tabular}{cc}
 \psfig{figure=bzcam_OIII.ps,width=8.6cm,%
     bbllx=4.1cm,bblly=8.1cm,bburx=17.7cm,bbury=26.05cm,clip=} &
 \psfig{figure=bzcam_Halpha.ps,width=8.6cm,%
     bbllx=4.1cm,bblly=8.1cm,bburx=17.7cm,bbury=26.05cm,clip=} \\
 \end{tabular}
 \caption[oiii]{The BZ Cam nebula in \ion{[O}{III]} $\lambda$5007 (left) and
        H$\alpha$ $\lambda$6563 (right), obtained with
        the WIYN telescope at 0\farcs8 seeing. The orientation of the slit 
        during the Tautenburg spectrographic observations,
        from which the line ratios have been derived (see Fig. \ref{lrat})
         is marked in the left frame. North is up and East to the left.
 \label{oiii}}
\end{figure*}

\subsection{The nebula revisited}

\begin{figure}[ht]
 \vbox{\psfig{figure=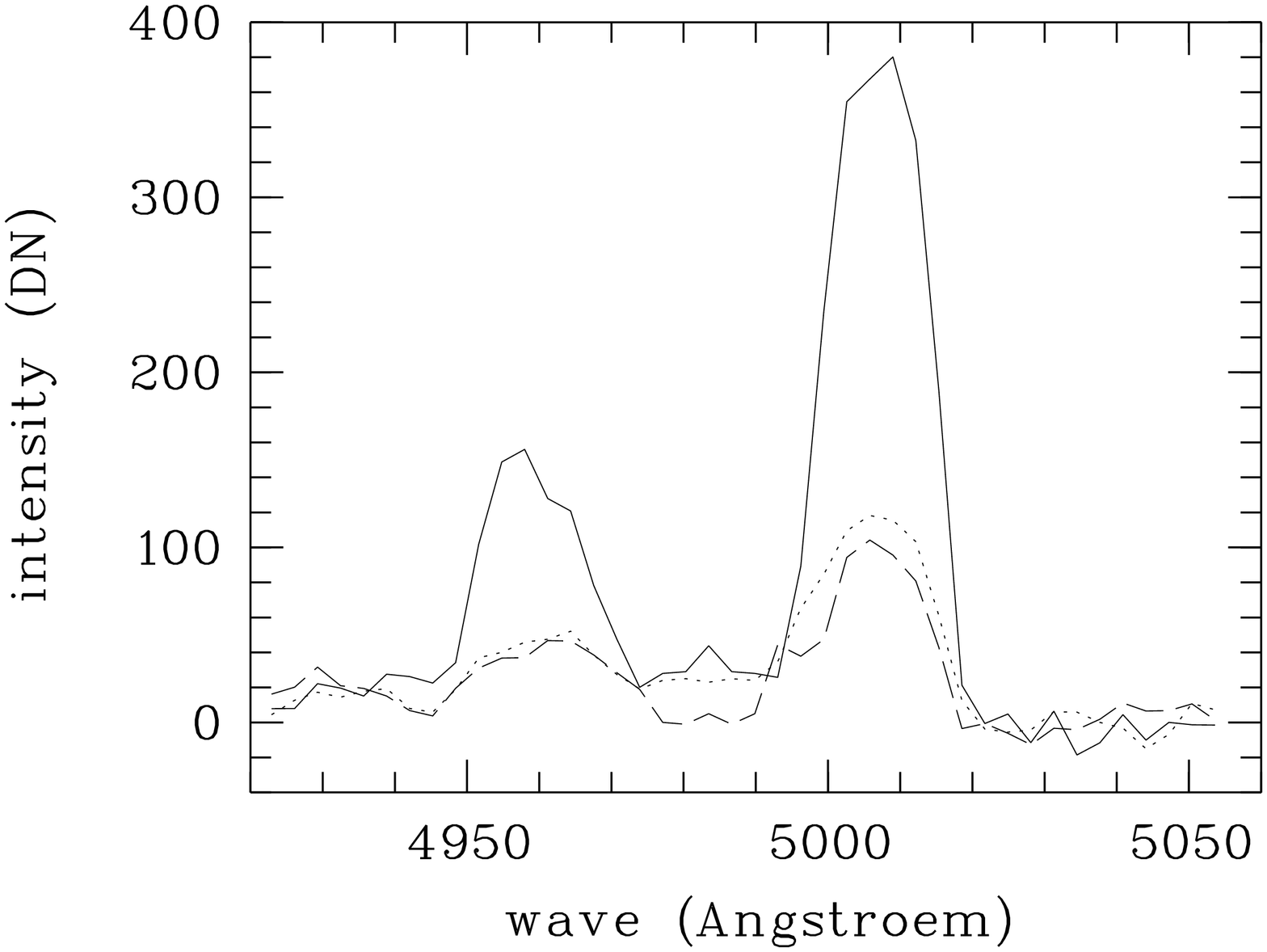,width=1.15\columnwidth,angle=270}}
 \vspace{-0.6cm}
 \caption[nebsp]{Spectrum of the BZ Cam nebula in the region of the
  \ion{[O}{III]} line, plotted for three different regions in the nebula:
  Region ``A'' (solid line), a region at the same position angle as ``A''
  but 4 times further away from BZ Cam (dashed line), and region
  ``B'' (dotted line).
 \label{nebspec} }
\end{figure}

\begin{figure}[ht]
 \vbox{\psfig{figure=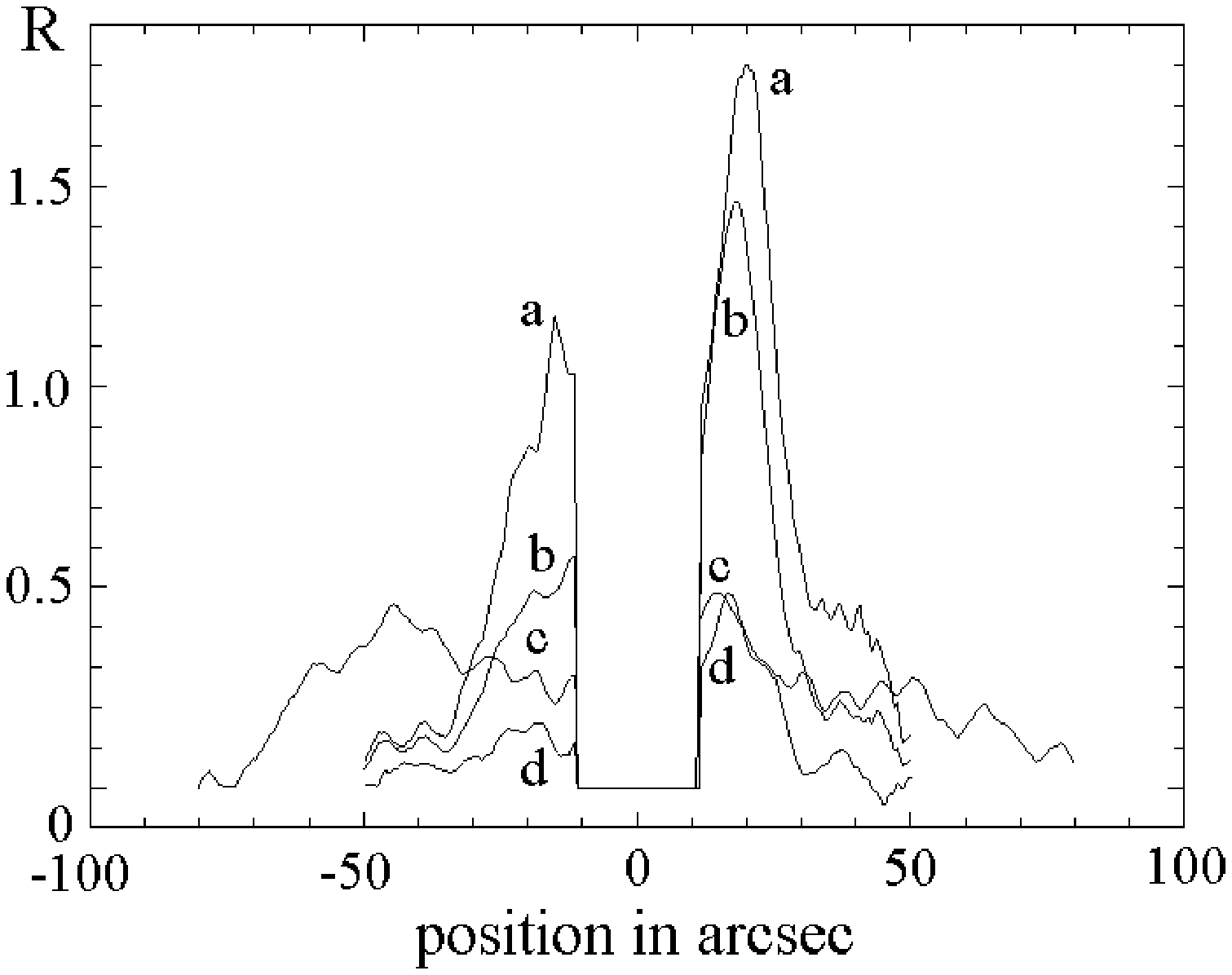,width=1.05\columnwidth}}
 \caption[lrat]{Relative line ratios $R$ across the BZ Cam nebula along the
  direction shown in Fig. \ref{oiii}. The central part has been omitted
  (BZ Cam itself).
  Negative values in the abscissa correspond to the North-western part of
  the slit in Fig. \ref{oiii}, positive values to the South-eastern part.
  The letters denote the following
  ratios: a: $R$ = 0.5*[OIII]/[NII]6583 \AA, b: $R$ = [OIII]5007 \AA/H$\alpha$,
          c: $R$ = [NII]6583 \AA/H$\alpha$,  d: $R$ = [OIII]4959 \AA/H$\alpha$.
 \label{lrat}}
\end{figure}

We also obtained narrow-band filter images in H$\alpha$ and the
\ion{[O}{III]} 4959/5007 \AA\ emission lines, showing the large nebula
around
BZ Cam (Krautter \etal\ 1987, Hollis \etal\ 1992).
Based on our higher sensitivity we find that the nebula seems to be
composed of two components: (i) a large, smooth component with bow-shock like
shape towards the South, and  a (ii) smaller, filamentary component
which also has a bow-shock like shape towards the south, but with a
smaller curvature radius, and in addition with cometary-tail like
extensions towards the North-East and North-West
(Fig. \ref{oiii}).
A nebula spectrum around the \ion{[O}{III]} line is displayed in
Fig. \ref{nebspec}.
In certain regions of the nebula the \ion{[O}{III]} line
is very strong relative to H$\alpha$, whereas in other regions it is not.
Fig. \ref{lrat} shows these
relative intensity variations for a few line ratios across the nebula.
One prominent example is the spectral difference of the two arcs marked as
``A'' and ``B'' in the right panel of Fig. \ref{oiii}. While both arcs
are similarly bright in H$\alpha$, only ``A'' is bright in
\ion{[O}{III]}.
This argues against shock-excitation, but is consistent with
photo-ionization from BZ Cam, since ``B'' is at a much larger distance
from BZ Cam.

We have estimated the total nebular flux in \ion{[O}{III]} to be
4.8$\times$10$^{-13}$ erg/cm$^2$/s (after a 16\% extinction correction
when using $E(B-V)$=0.05 from Verbunt 1987).
At a distance of 830 pc (Ringwald \& Naylor 1998) this corresponds to
$L_{\rm [OIII]} \sim\ 4\times 10^{31}$ erg/s.

\begin{table}
\caption{Relative optical emission-line fluxes of the nebula ($\sim$20\% error)
 at three different locations, compared to the Rappaport \etal\ 1994 
 prediction (last column).}
\vspace{-0.2cm}
\begin{center}
\begin{tabular}{lcccc}
   \hline
   \noalign{\smallskip}
   Spectral Line   & \multicolumn{3}{c}{Relative Flux Range in} & predicted  \\
                   &   app1       &  app2  & app3   &  \\
      \noalign{\smallskip}
      \hline
      \noalign{\smallskip}
 H$\beta$ & 0.29 & 0.37 & 0.55  & 0.28-0.37 \\
 \ion{[O}{III]}$\lambda4959$  & -- & 0.54 & 1.08 & 0.12-2.13 \\
 \ion{[O}{III]}$\lambda5007$  & -- & 1.53 & 3.01 & 0.25-6.15 \\
 H$\alpha+$N{\small{II}}$\lambda6548$  & 1.0 & 1.0 & 1.0 & 1.0 \\
 N{\small{II}}$\lambda6584$  & 0.46 & 0.37  & 0.55 &  0.43-2.5\\
 S{\small{II}}$\lambda$6717+6731$\!\!$  & 0.47 & 0.35 & 0.29 & 0.26-0.51 \\
 \noalign{\smallskip}
 \hline
 \end{tabular}
 \end{center}
 \label{linneb}
\end{table}

We have earlier speculated that based on the behaviour of the SSB
\rxj0513\ and the VY Scl star \v7\ possibly also other VY Scl stars
could be emitters of supersoft X-ray emission during the optical
low-state (Greiner \etal\ 1999).
In the case of BZ Cam, the similarity in the wind properties with V\,Sge
adds even more support to this conjecture. We will argue in the following
that by assuming supersoft X-ray emission during the optical low-states
we can explain the hitherto puzzling properties of BZ Cam's nebula.

We first note that
all measured nebular emission line ratios ([OIII], [NII], [SII] vs.
H$\beta$) of the nebulae surrounding BZ\,Cam (our own, Tab. \ref{linneb}, 
as well as those of Hollis \etal\ 1992) are in agreement with predictions
for supersoft X-ray source nebulae (Rappaport \etal\ 1994).
Second, models of such ionization nebulae
(Rappaport \etal\ 1994) show that about 2--8\% of the total
ionizing flux is re-radiated by \ion{[O}{III]}. These models have been
calculated for luminosities of 10$^{37}-10^{38}$ erg/s and densities
of 1--12 cm$^{-3}$. In the case of BZ Cam, the ionizing luminosity
will potentially be smaller (see below), and the density substantially
higher: Krautter \etal\ (1987) estimates electron densities of 
$n_{\rm e}$ $\sim$ 100--250 cm$^{-3}$ under the assumption of an 
electron temperature of 10000 K.
This implies that a model more appropriate for BZ Cam could be expected to
re-emit substantially less than 2\% of its X-ray flux in \ion{[O}{III]}.

To make this expectation somewhat more quantitative, we used the XSTAR 2.0
code (Kallman 2000) as distributed in the HEAsoft package, and computed
the emission line luminosities over a grid of the following input parameters:
(i) hydrogen density $n$ = 35 and 100 cm$^{-3}$;
(ii) temperature of the central, ionizing source $T$ = 15 and 25 eV.
The following parameters were adopted and not varied:
luminosity of the central source $L_{rm X}$ = 10$^{36}$ erg/s;
outer radius (in \ion{[O}{III]}) of the nebula $r_{\rm max}$ = 0.6 pc.
We find that the luminosity of the \ion{[O}{III]} $\lambda$5007 line
is in the range of 0.1-2\% of the X-ray luminosity.
Thus, based on our measured \ion{[O}{III]} flux,
the mean  (over the last 10000 yrs) ionizing (X-ray) luminosity would be
L$_{\rm X}^{\rm mean}$\,$\sim$ 2$\times$10$^{33}$--4$\times$10$^{34}$\,erg/s.

Only 2 optical low-states of BZ Cam are known over the last 110 years,
separated by 71\,yrs. As mentioned in the introduction, the duration of the
1999/200 optical low state is difficult to determine, but is shorter than
280 days. The duration of the 1928 low-state was less than about 100 days.
Assuming 180 days as a mean low-state duration, results in a duty cycle
of about 1/150.
Thus, if supersoft X-ray emission occurs only
during the optical low-states (as e.g. in \rxj0513), then we can deduce
an ionizing luminosity (for \ion{[O}{III]}) during the optical low-state
of L$_{\rm X} \sim$ 3$\times$10$^{35}$--6$\times$10$^{36}$ erg/s.
Such an X-ray luminosity is very similar to that observed from
the VY Scl star V751 Cyg (Greiner \etal\ 1999)
or V Sge (Greiner \& Teeseling 1998) during their optical low-states.
Note that \ion{[O}{III]} is only ionized if the effective temperature of the
white dwarf is hot enough, say $\gax$10 eV, which is expected to primarily
happen during optical low-states when X-ray emission is on. In contrast,
hydrogen requires a lower ionizing potential, and therefore can be expected
to be ionized also during the optical high states.

Since the size of the ionization zone is proportional to the ionizing flux
(Rappaport \etal\ 1994), a rough estimate of the expected size of BZ Cam's 
nebula can be made by comparison to the canonical SSB CAL 83 and its nebula
(Remillard \etal\ 1995). With our X-ray 
flux, the expected size of the ionization nebula (in H$\alpha$) of
BZ Cam should be a factor 50 smaller than that of CAL 83, i.e. 0.4 pc. 
Indeed, the size of the BZ Cam
nebula in H$\alpha$ is 2\farcm5 $\times$ 5\amin, corresponding to
0.6 $\times$ 1.2 pc at 830 pc distance.
Thus, the East-West extension of the nebula (perpendicular to the proper 
motion) is in perfect agreement with the prediction for an ionization nebula.

With the knowledge of the proper motion of BZ Cam,
the bow-shock-like shape of the nebula as well as the two-component
appearance can be explained based on a moving ionizing source,
very similar to the theoretical considerations of Chiang \& Rappaport
(1996).
Note that the time scale for illumination/photo-ionization
is generally smaller than the time scale for recombination.
In addition, the ionization time scale is fairly insensitive to the
source luminosity (during the on-state) so that we can use the analogy
to the more luminous, canonical 10$^{37-38}$ erg/s SSBs 
(Chiang \& Rappaport 1996).
As a source emits ionizing photons, the ionization front propagates
radially outward and decelerates as the ionizing flux attenuates
due to both geometric (1/r$^2$) dilution and photoelectric absorption.
If a source moves substantially before the ionization front
nears its equilibrium radius, the ionization nebula will become
elongated,
i.e. while ahead of the motion fresh gas is continually ionized,
a slowly fading wake of recombining ions is left behind.
Using formula (7) of Chiang \& Rappaport (1996), and using
$\rho \sim 35$ cm$^{-3}$ and L$_{\rm X}^{\rm mean}$, this will
happen for BZ Cam at any velocity larger than 50 km/s.
With our above derived space velocity of 125 km/s (assuming  830 pc
distance) and a homogeneous surrounding interstellar medium one theoretically
would expect
an axis ratio of about 1.4 (see Fig. 8 in Chiang \& Rappaport 1996).
Again, this is in surprising agreement with the value for BZ Cam of 1.3
as measured from the H$\alpha$ image (right panel of Fig. \ref{oiii}).

As shown earlier (Fig. \ref{oiii}), BZ Cam's nebula seems to consist of
two components. We believe that the above description, i.e. ionization
by a moving source, applies to both
components. However, the smaller components with the filaments
(which are very bright in the line emission)
may represent regions of higher density which could be shaped by both
the ram pressure of the bow-shock (ahead of BZ Cam's motion) as well as
episodes of transient major ejection events (in the far tail of
BZ Cam's nebula). Both the larger size of low-excitation emission
as compared to e.g. \ion{[O}{III]} as well as the distance dependence
of the emissivity in \ion{[O}{III]} argue in favor of photo-ionization
also of the small, filamentary component which is just shaped, but not
excited by the hydrodynamic shock formed by BZ Cam's motion.

\section{Conclusions}

Our conclusions can be summarized as follows:
\begin{itemize}

\item
We have
(1) discovered proper motion of BZ Cam, leading to a space velocity of 
125 km/s;
(2) measured variable \ion{[O}{III]}/H$\beta$ ratios depending on the
  distance to the ionizing binary;
(3) noted that the emission line ratios of BZ Cam's nebula are in agreement to
   predictions of ionization by supersoft X-ray emission;
(4) measured the \ion{[O}{III]} flux which under the assumption of
  being recombination radiation and the knowledge of the duty cycle
  of optical low and high states of BZ Cam implies a mean ionizing soft
  X-ray flux of 3$\times$10$^{35}$--6$\times$10$^{36}$ erg/s;
(5) shown that the transversal size of BZ Cam's nebula is in agreement to the
  prediction of ionization by supersoft X-ray emission as measured from
  the \ion{[O}{III]} flux, i.e. the size is compatible with the nebular
  \ion{[O}{III]} flux;
(6) explained the North-South extent of the nebula by a moving source,
  and found that the axis ratio of the nebula implies a velocity which
  is in agreement with the measured space velocity;
(7) argued that only the shape of the small, filamentary component of the
  nebula could be a ram-pressure formed bow-shock, while the excitation
  of both nebular components is due to ionization.
\newline
This naturally explains the nebular emission line ratios, the shape, size
and flux of BZ Cam's nebula, and
avoids the many complications related to a shock-excitation
interpretation as proposed by Krautter \etal\ (1987), and elaborated by 
Hollis \etal\ (1992), which among others include
(1) the radio emission is thermal,
(2) the radio nebula anticorrelates with [OIII],
(3) the large [OIII]/H$\beta$ ratio,
(4) and the necessity to truncate the recombination zone
by e.g. assuming  an environment with larger rather than lower density.

\item
The deduced mean X-ray ionizing flux indeed suggests that BZ Cam
is a transient supersoft source, presumably during optical low states.
This supports our earlier conjecture that the whole class of VY Scl stars,
or at least many of them, could be transient supersoft sources 
(Greiner \etal\ 1999). The small \ion{He}{II} emission line strengths
in BZ Cam (and similarly in V751 Cyg) can be understood if VY Scl stars
are the low-mass extension of the canonical SSB with concordantly
lower effective temperatures and hence less photons which are capable of
ionizing \ion{He}{II}.
Like other VY Scl stars, BZ Cam is a known, hard X-ray source during optical
high state (van Teeseling \& Verbunt 1994, Greiner 1998), 
noticable similar to V Sge and V751 Cyg. The origin of this 
hard X-ray emission is presently not clear.

\item
The P Cyg profiles of emission lines (wind with few thousand km/s)
are absent during the optical low-state. 
At present, it is not clear how this wind
is related to the phase of conjectured supersoft X-ray emission.
Note, however that for orbital periods shorter than 4--5 hrs, as is the
case for BZ Cam, the canonical interpretation of SSBs with thermal
timescale mass transfer from a companion more massive than the white
dwarf is not applicable. Instead, large mass overflow is possible in
wind driven systems (van Teeseling \& King 1998). BZ Cam may be the
first system where we observe such a wind.

\item
The ultimate proof of our scenario is simple: detect luminous, supersoft
X-ray emission during the next optical low state of BZ Cam.

\end{itemize}

\begin{acknowledgements}
Much of the optical data presented in Fig. \ref{lc} were taken from the
VSNET which we kindly acknowledge.
This research has made use of the SIMBAD database and the
VizieR Catalogue Service, Strasbourg, of the
Digitized Sky Survey data produced at the Space
Telescope Science Institute, Baltimore and of the
ESO Skycat Tool, version 2.1.1.
RDS gratefully acknowledges financial support from the Deutsches Zentrum
f\"ur Luft- und Raumfahrt (DLR) (F\"or\-der\-kenn\-zeichen 50~OI~0001).
We thank the referee, K. Mukai, for constructive comments.
\end{acknowledgements}

\end{document}